\documentclass[preprint,showpacs,preprintnumbers,amsmath,amssymb]{revtex4}
\usepackage{graphicx}
\begin{document}
\title{Some aspects on four quarks recombination}

\author{G. Toledo S\'anchez and I. A. Toledano Ju\'arez }

\affiliation{ Instituto de F\'{\i}sica, Universidad Nacional Aut\'onoma de M\'exico. M\'exico D. F. C.P. 04510}
\date{\today}

\begin{abstract}

We have performed a 3-D Monte Carlo simulation of a system composed of two identical light quarks ($qq$) and two identical antiquarks ($\bar Q\bar Q$) and 
 determined whether it is energetically more favorable to form a tetraquark or two mesons, as a function of the interparticle separation distance which, for a fixed number of particles, can be identified as a particle density.
 In this proceedings, we highlight the main results and elaborate on the implications in properties like  the correlation function for two-mesons and characterize the isolated diquark correlation function.  We analize the four-body potential evolution and exhibit its linear behavior as a function of the invariant distance. We track the dynamical flipping among configurations to determine the recombination probability, exhibiting the importance of the tetraquark state.
 \end{abstract}


\maketitle

\section{Introduction}

 Recent experimental research has provided strong evidence on the formation of tetraquark states \cite{Belle,Bes,lhcb,dzero}. A new era in hadron spectroscopy is just ahead \cite{sakai}, calling for a better understanding of the strong interaction in a scenario where conventional mesonic and baryonic systems try to cope with the experimental data. The early theoretical studies inquired on the existence and stability of the tetraquark\cite{Jaffe,isgur85,lenz,carlson,stancu94,Val,CFT,santopinto,reviewIJMPA}, and how its mixing with a meson state can help us to understand the observed spectroscopy of states like the $\sigma$ meson \cite{reviewIJMPA, mixing}.  
 
A tetraquark state offers the opportunity to learn about the strong interaction features leading to its formation when two mesons are forced to approach each other as it could happen in a meson-meson collision or, on the opposite, when the four quarks are produced very close in space as in the $WW$ decay, which eventually freeze out to two mesons \cite{wmass,ww}. 
In these scenarios several questions can be asked:
i) Under which conditions two mesons can turn into a tetraquark or mixed state? ii) Is there any   property at the meson level that can reflect the presence of an intermediate tetraquark state?
iii) What is the structure of a diquark? iv)How the many-body potential behaves? v) Can we turn it into a simpler form? vi) Can we distinguish between a molecular (meson-meson bound) and a  tetraquark state? vii) Is the quark recombination relevant for hadronic states? viii) How is it modified by the presence of the tetraquark?, etc.  \\
In a recent work \cite{ivan15}, we addressed these questions using an effective model to mimic the strong interaction among quarks. In this proceedings we highlight the main results and elaborate on the implications. Let us recall the procedure followed:

 We consider two identical light quarks $qq$ (denoted by $u$) and two identical antiquarks $\bar Q\bar Q$ (denoted by $d,s,c$ and $b$, depending on the constituent mass with respect to the light  one). The seminal work by Lenz {\it et al.} \cite{lenz} laid down a procedure to describe such a system in this context and found its general properties, as isolated objects, using a harmonic potential interaction. We perform a 3-D Monte Carlo simulation of the system including quantum correlations between particles to determine whether it is energetically more favorable to form a 4-body state (tetraquark) or two mesons, as a function of the interparticle separation which, for a fixed number of particles, can be identified as a particle density. We are interested in s-wave states where tensor and spin interaction effects are expected to be negligible in the gross features of the properties.

\section{Meson variational wave function}
Let us start with the description of the exact solution and the variational approach of a meson state, composed of a quark and an antiquark of mass $m_1$ and $m_2$ respectively. The strong interaction between this pair can be represented by a flux tube \cite{isgur85}, which can be effectively described by a linear potential $V \left[ \vec{r}_1, \vec{r}_2 \right] = k \left\vert \vec{r}_1 - \vec{r}_2 \right\vert = k r$ where $k$ is an interaction constant and $r=|\vec{r}_1 -\vec{r}_2|$ is the relative distance between them. Lattice QCD studies have confirmed the linear behavior of the interaction among quarks at large distances \cite{qqlinear}.
The exact solution corresponds to the eigen-energies and eigen-functions:
\begin{equation}
E_n = \left[ \frac{k^2}{2\mu} \right]^{1/3} |\xi_n|, \hspace*{1cm} \rho_n (r) = \frac{1}{r} \mbox{Ai} \left( r \left[ 2 \mu k \right]^{1/3} - \xi_n \right),
\label{EAiry3D}
\end{equation}
where $\xi_n$ are the zeros of the Airy function ($ \mbox{Ai}$) and $\mu$ is the reduced mass. The ground state corresponds to the first zero of the Airy function, with energy $E_0 = 2.3381$,
 in $m_1 = m_2 = m = k = 1$ units.
A variational wave function that approaches the exact result for the ground state takes the following form:
\begin{equation}
F_\lambda (r) = \sqrt{\frac{3 \lambda^2}{2 \pi}} e^{-\lambda r^{3/2}},
\label{FlambdaMeson}  
\end{equation}

\noindent where  $\lambda$ is the variational parameter and $r$ is the relative coordinate of the two-body system. We have found that the wave function reproduces the exact minimum energy value within a 1\% error. This simple form allows to find analytical expressions for the expected variational parameter, energy and mean square radius. By matching them to their experimental values we can fix the physical units. 

\section{Four-body system}
Using the previous results, we can now extend our study to a four-body system composed of two identical $u$-like quarks ($qq$) and two identical antiquarks ($\bar Q \bar Q$), which  can be $d,s,c$ and $b$-like. We will refer to this four-body system by $qQ$.
At very low density, where the distance between quarks is large, the model must reproduce a system of two isolated mesons. In addition, as the density increases, the inter-particle separation becomes small allowing the interaction between the two mesons, which can be represented by quark exchange or a truly four-body interaction. All over, the Pauli blocking among identical particles must be enforced. We rely on the String-Flip Model \cite{lenz,moniz,oka} which has been also used to study dense matter systems \cite{Piekarewicz,toledo02,ayala}, to describe the dynamical transition among these regimes.

\begin{figure}
\begin{center}
\includegraphics[width=17pc]{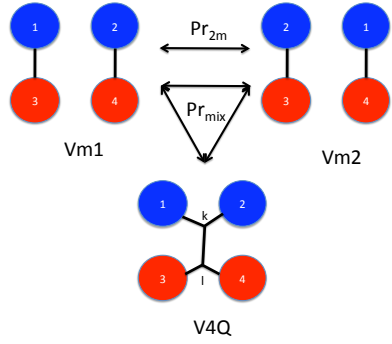}
\caption{\label{4potential} Potential configurations for the system (Vm1,Vm2,V4Q). 1 and 2 are the particles and 3 and 4 the antiparticles. Pr$_{2m}$ and Pr$_{mix}$ are the possible quark recombinations for the two mesons and mixed system, respectively.}
\end{center}
\end{figure}

The strong interaction of a many-body system can be well approached by considering that the quarks are connected by gluon flux tubes  \cite{isgur85} according to a configuration producing the lowest energy state of the whole system.
Let us denote the position of the particles by $\vec{r}_1$ and  $\vec{r}_2$ and the antiparticles by $\vec{r}_3$ and $\vec{r}_4$. The way to link the four particles of the system consistent with the QCD restrictions of color neutrality are shown in Figure \ref{4potential}:
 i) The system is composed of two mesons ($V_{m1}$ or $V_{m2}$). ii) The system is composed of a tetraquark ($V_{4Q}$). In this case, the shortest path linking the four particles is given by the Steiner-tree, which uses two auxiliary vectors (denoted by $\vec{k}$ and $\vec{l}$) to minimize the length of the configuration. The vector $\vec{k}$ links the diquark sub-system and the vector $\vec{l}$ the anti-diquark one. 
When all these configurations are allowed, the potential of the system is chosen as the one producing the minimum potential energy in a given configuration:
\begin{equation}
V = min (V_{m1}, V_{m2},V_{4Q}).
\end{equation}

We propose the following variational wave function, which reproduces the two isolated mesons at very low densities, allows four-body correlations and incorporates the Fermi correlations as the density increases:
\begin{equation}
\Psi_{\lambda} = \Phi_{FG} \left( e^{-\lambda Q} \right),
\label{WFgeneral}
\end{equation}

\noindent where $\lambda$ is the single variational parameter, $Q$ is a function driven by the many-body potential, and $\Phi_{FG}$ is the free Fermi system wave function.
For a four-body system, the $Q$ function takes the following form, depending on the optimal potential at the given configuration, denoted by the sub-index:
\begin{equation}
Q_{m_1} =  r_{13} ^{3/2} +r_{24}^{3/2} ,
\hspace*{1cm}
Q_{m_2} =  r_{14}^{3/2} + r_{23}^{3/2} 
\end{equation}
or
\begin{equation}
Q_{4Q} =  r_{1k}^{3/2} + r_{2k}^{3/2} + r_{kl}^{3/2} + r_{3l}^{3/2} + r_{4l}^{3/2} ,
\end{equation}
where the 3/2 power resembles the variational solution of single pairs linked by a linear potential, described in the previous section.
$\Phi_{FG}$ is composed of single wave functions of a particle in a cubic box of side $L$. We can define a particle density parameter as a measure of the interparticle separation by:
$\rho = N/V= 4/L^3$, where $N$ is the number of particles of the system and $V$ is the box volume. For a fixed number of particles, the change in the particle density corresponds to modify the box size and correspondingly to the inter-particle separation.

\begin{figure}
\begin{center}
\begin{minipage}{16pc}
\includegraphics[width=16pc]{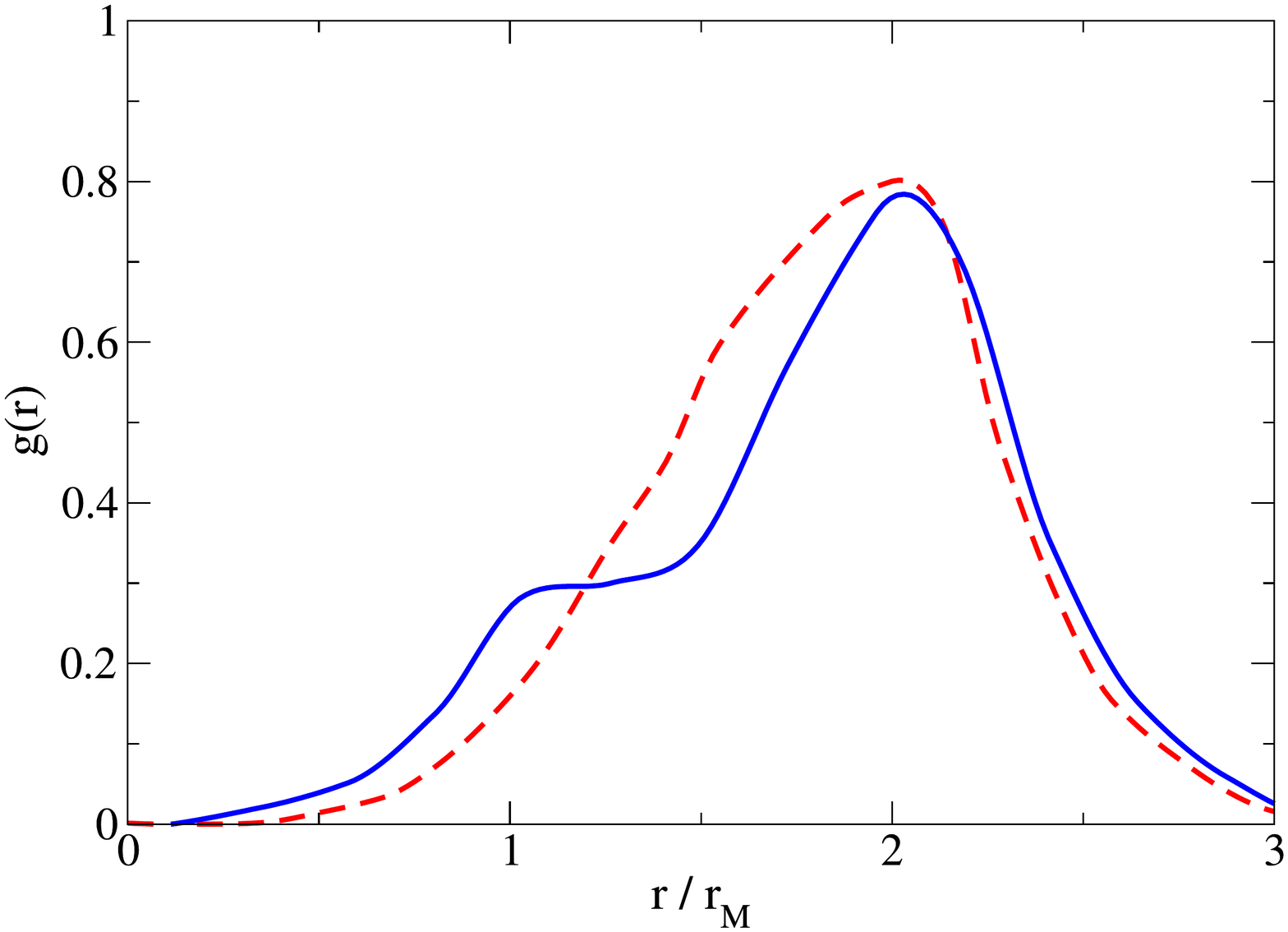}
\caption{\label{corre2m} Two meson correlation function for $ud$ at $\rho=0.01$. Dashed (Solid)  line corresponds to the two mesons (mixed) case. }
\end{minipage}\hspace{2pc}%
\begin{minipage}{16pc}
\includegraphics[width=17pc]{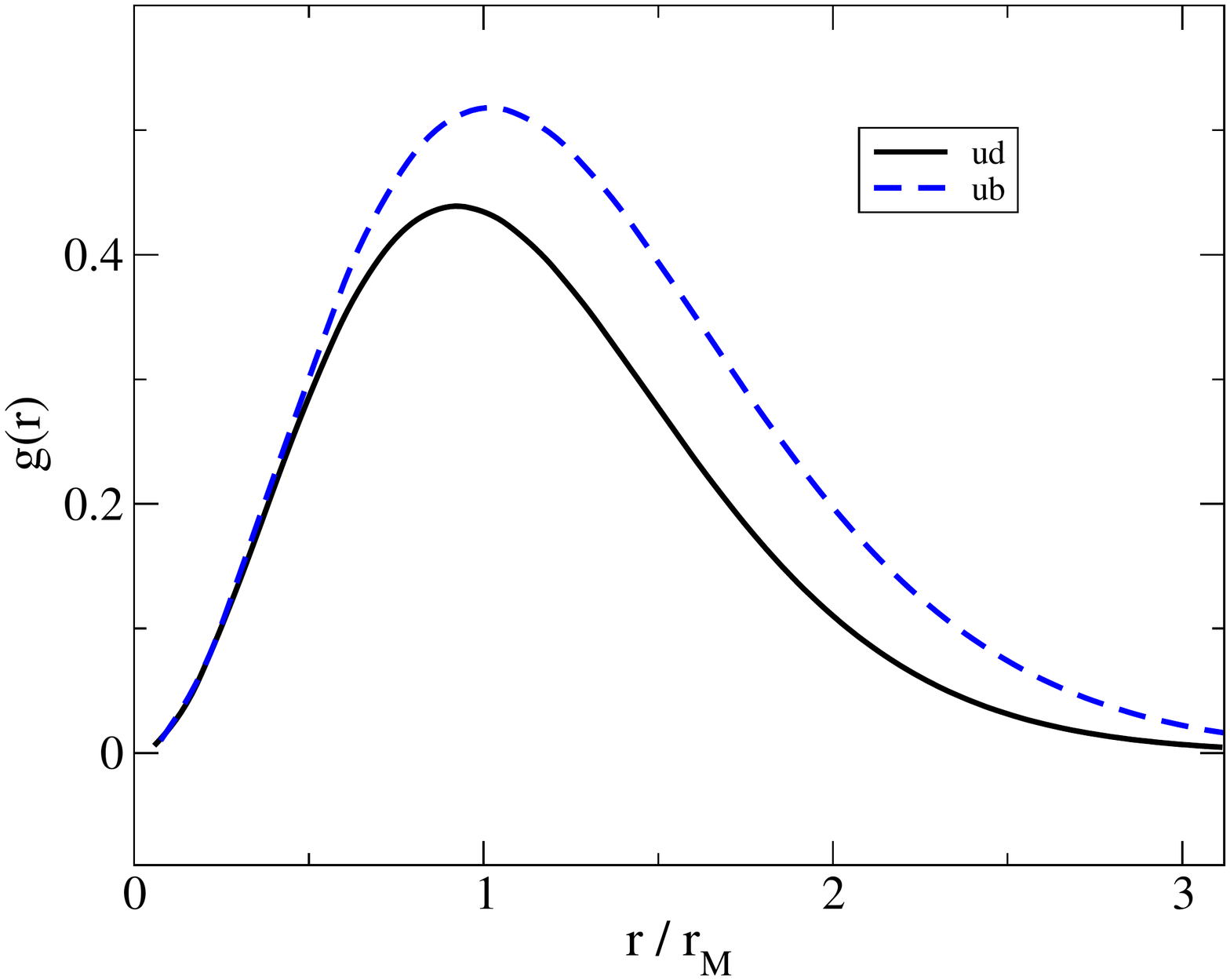}
\caption{\label{diquarkgr}Diquark correlation function at zero density.}
\end{minipage} 
\end{center}
\end{figure}

\section{Results}
The expectation value for the energy is computed using Monte Carlo techniques, in particular we use a Metropolis algorithm for the sampling, driven by the square of the wave function.
 We computed the properties of three systems:

\begin{itemize}
 \item \emph{Two mesons}. In this case, only meson configurations are allowed in the potential. We characterize the meson properties, comparing the results with the expected from the analytical solution in the isolated limit and the departure from them as density increases. 
 \item \emph{Tetraquark}.  In this case, only tetraquark configurations are allowed in the potential. We characterize the tetraquark properties in the isolated limit and the departure from it as density increases.
 \item \emph{Mixed}.  In this case, all the possible configurations are allowed in the potential. We characterize the meson and tetraquark properties and their modifications due to the presence of the other configuration.
\end{itemize}
For all the above cases we explored the mass effect by considering the two quarks to be light ($u$)and the antiquarks having variable mass to resemble the  $m_{d,s,c,b}/m_u$ ratios.
Results are presented for several light-quarks to heavy-antiquarks mass ratios whenever  they are found to be relevant.

\subsection{Correlation functions}
A useful observable to characterize the properties of the system is the two-particle correlation function. It measures the probability of finding two particles at a relative distance $r$ from each other. In Figure \ref{corre2m}, we show the meson-meson correlation function at a density $\rho=0.01$ for the two-mesons (dashed line)  and mixed case (solid line). We exhibit the results corresponding to a light-light ($ud$) system but a similar behavior is observed in the other mass ratio combinations. The comparison between both figures shows that there is a modification in the meson-meson correlation function by the presence of the tetraquark state at intermediate densities, a bump develops in the near tail of the correlation function, driven by the diquark formation and fades out as the density increases and all the quark correlations vanish. This may be an alternative observable to have indirect evidence of the tetraquark formation.

For the tetraquark case, we can explore the structure of the quark-quark (diquark) correlation function. In particular, in the zero density limit, the normalized tetraquark system correlation function for the diquark  behaves as shown in Figure \ref{diquarkgr} for the light-light (solid line) and light-heavy (dashed line) cases, which can be parameterized by:
\begin{equation}
g(r)_{diquark}= A_0 r^2 e^{-r^{A_2}/A_1^2}.
\end{equation}
In Table \ref{diquarkpar}, we show the value of the parameters for each case. Note that the dependence on $r$  of  the diquark correlation function is close to  others suggested in the literature \cite{formfactor,bicudo13} but in our case it is obtained in a dynamical way. The short distance region shows the competition between the Pauli blocking among the identical particles and the attractive potential.
\begin{table}
\caption{\label{diquarkpar}Diquark  correlation functions parameters for $ud$ and $ub$-like mass ratios.}
\begin{center}
\begin{tabular}{llll }
\hline
System & $A_0$ & $A_1$ &$A_2$\\ 
\hline
 ud &
 0.64&
 1.24&
 1.51\\
 ub &  
 1.13&
 1.1&
1.47    \\
 \hline
\end{tabular}
\end{center}
\end{table}

\subsection{Tetraquark potential}
\begin{figure}
\begin{center}
\begin{minipage}{17pc}
\includegraphics[width=17pc]{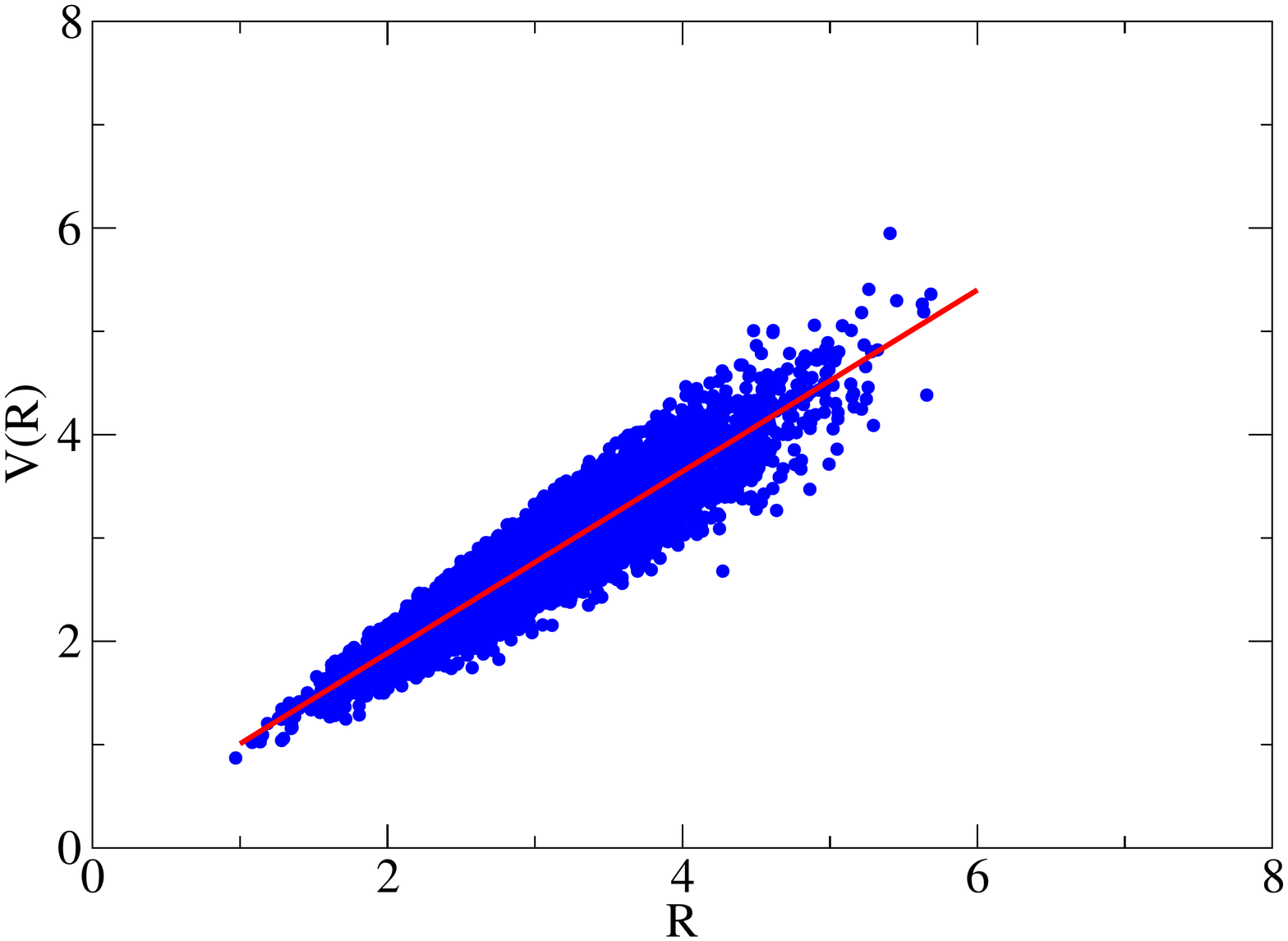}
\caption{\label{linearpotential} Tetraquark potential obtained in the simulation (symbols) as a function of $R$. The solid line is a linear fit to the data.}
\end{minipage}
\hspace{2pc}%
\begin{minipage}{17pc}
\includegraphics[width=16pc]{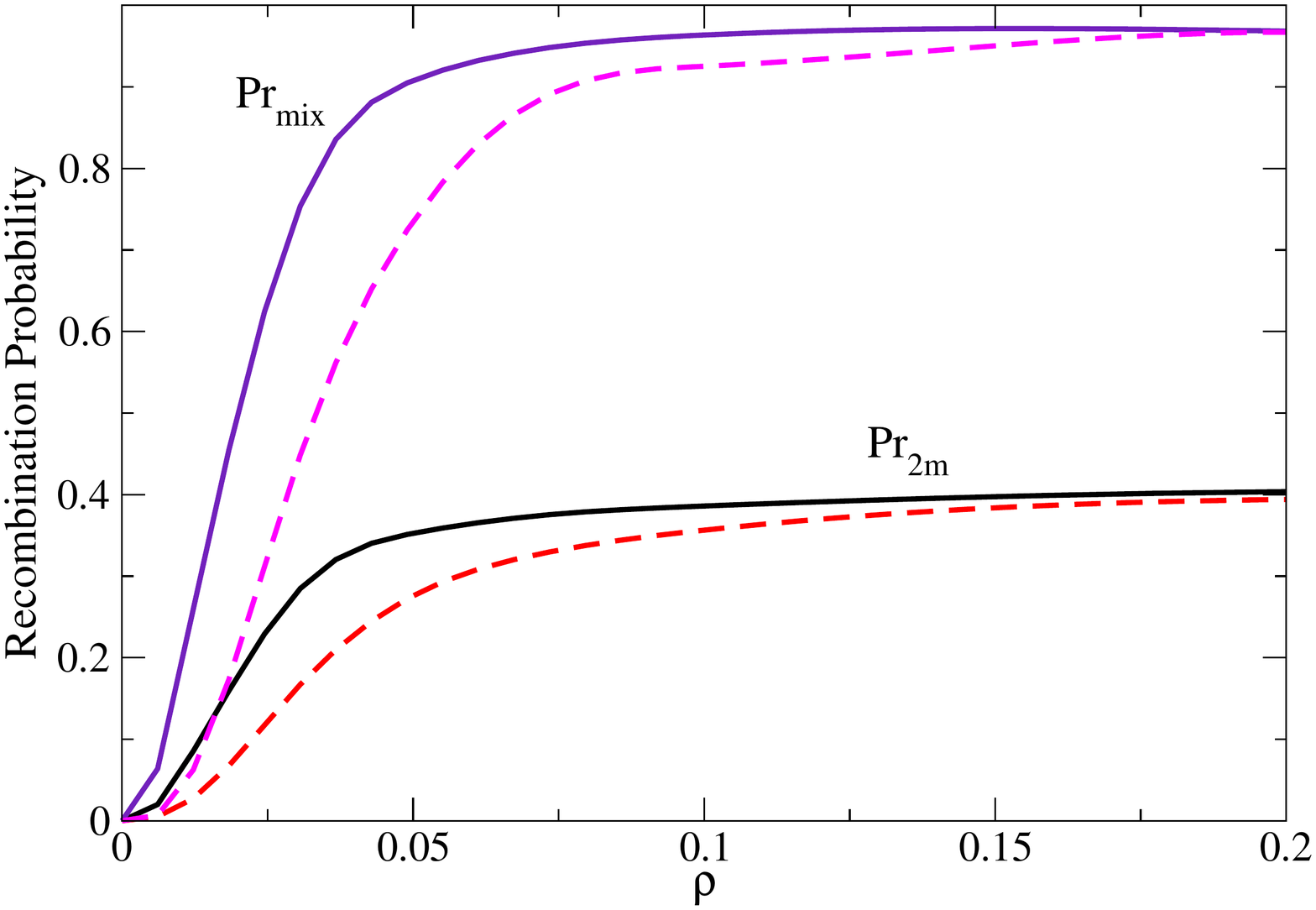}
\caption{\label{recombination}Recombination probabilities as a function of the density, for  two mesons ($Pr_{2m}$) and mixed ($Pr_{mix}$) systems. The solid  and dashed lines are for $ud$  and $ub$ systems respectively.}
\end{minipage} 
\end{center}
\end{figure}

The tetraquark potential depends on the quarks positions and two auxiliary vectors, placed in such a way that the total length linking the quarks is the shortest one.  These auxiliary vectors are modified in a non-trivial way whenever a single quark changes its position, making this a truly many-body interaction. An effective behavior of the potential can be set as linear respect to the invariant length, $R\equiv \sqrt{\sum r^2_{ij}}$:
\begin{equation}
V(R)=R_0+BR,
\end{equation}
where $R_0$ is the value of the potential at zero distance, which is expected to be modified by the short distance coulomb-like correction. In Figure \ref{linearpotential}, we show the tetraquark potential values as obtained from the simulation (symbols) as a function of $R$ and a linear fit to them (solid line) when only tetraquarks are allowed. The simulation data clearly follows the linear fit, The slope shows small dependence on the density, and $B(\rho \approx 0) =0.84\pm 0.02$. In the mixed case there is a significant density dependence, starting below  the corresponding value at zero density ($B(\rho \approx 0)=0.67\pm 0.02$) and as the density increases they approach to each other. This behavior is similar when considering different flavor  systems.

An effective four-body contact potential, $V_4(contact)$,  as a function of density, can be determined as the average potential irrespective of the length linking the particles.
\subsection{Dynamical recombination}
The  four quark recombination can have important effects in systems where they are produced very close in space. An example can be found in the $WW \to qq\bar Q \bar Q$ decay, whose spatial separation at LEP2 energies is around 0.1 fm \cite{wmass}. Although in the perturbative regime the recombination is small, in the non-perturbative regime the effect may be important.
The typical scenarios to estimate the recombination are: Considering spherical or elongated bags color sources and the reconnection is proportional to the overlap of two color sources;
strings considered as vortex lines where reconnection takes place when the cores of the two string pieces cross each other \cite{ww}. In our case, we would have, in addition, recombinations similar to tetraquark states which eventually freeze out to two mesons. The flipping from one configuration to another, driven by the minimal potential energy restriction, is a measure of the dynamical recombination property of the system. We define the recombination probability for the two mesons case  by $Pr_{2m}$, denoting the number of flippings among meson configurations normalized to the total configurations. Considering the tetraquark configuration as contributing to the recombination, the probability $Pr_{mix}$ counts the flipping among all of them, as depicted in Figure \ref{4potential}.
The behavior of these probabilities as a function of density are shown in Figure \ref{recombination}. We observe a difference between them in size, which can be attributed to the normalization, and more subtle in shape, as the rising to the high density convergence value is faster approached in the mixed case.\\ 

We have presented some general features about a four quarks system recombination states but
many questions are still open. The  inclusion of more properties and refinement in the approaches are certainly important to be considered in future work.
 
\begin {acknowledgments}
We congratulate the Mexican Division of Particles and Fields for its $30^{th}$ annual meeting, and thank the organizers for a very pleasant gathering.
\end {acknowledgments}


\begin{thebibliography}{100}
 
\bibitem{Belle} Z.Q. Liu et al. (Belle Collaboration), Phys. Rev. Lett. {\bf110}, 252002 (2013).
\bibitem{Bes} M. Ablikim et al. (BES III Collaboration), Phys. Rev. Lett. {\bf110}, 252001 (2013).   
\bibitem{lhcb}    LHCb Collab. (R. Aaij et al.), Phys. Rev. Lett. {\bf112}, 222002 (2014).
\bibitem{dzero}V. M. Abazov, et al., (Dzero Collab.), Phys. Rev. Lett. 117, 022003 (2016)
 \bibitem{sakai} A.~Hosaka, et al., PTEP 2016 (2016) no.6, 062C01

 \bibitem{Jaffe} R. L. Jaffe,  Phys. Rev. D {\bf 15} 267(1977).
\bibitem{isgur85} N.~Isgur and J.~Paton  Phys.~Rev.~D {\bf 31} 2910(1985).
\bibitem{lenz}  F.~Lenz et al.,  Annals of Phys. {\bf170} 65(1986). 
\bibitem{carlson} J.~Carlson and V.~R.~Pandharipande Phys. Rev. D {\bf 43} 1652(1991).
\bibitem{stancu94}D.~M.~Brink and F.~Stancu, Phys.~Rev.~D {\bf 49} 4665(1994). 
\bibitem{Val} J.~Vijande, A.~Valcarce, J.~M.~ Richard, Phys. Rev. D {\bf 87}
  034040(2013).
\bibitem{CFT} Pedro Bicudo, Nuno Cardoso, Marco Cardoso. Prog.~Part.~Nucl.~Phys.~{\bf67} (2012). 440 
\bibitem{santopinto} J. Ferretti, G.~Galat\`a and E.~Santopinto Phys. Rev. D {\bf 90} 054010(2014).

\bibitem{reviewIJMPA} A. Esposito et al, Int.~Jour.~Mod.~Phys.~A, {\bf 30} 1530002(2015).

\bibitem{mixing} Ping Wang, Stephen R.~Cotanch and Ignacio J.~General, Eur.~Phys.~ J.~ C. {\bf 55} 409(2008)
\bibitem{wmass} G. Abbiendi et al, (The OPAL Collaboration),  Eur.~Phys.~ J.~ C. {\bf 45} 307(2006). 
\bibitem{ww} T. Sj\"ostrand and V. A. Khoze, Z.~Phys.C {\bf 62} 281(1994); Phys. Rev. Lett. {\bf 72} 28(1994). 
\bibitem{ivan15} I. A. Toledano Ju\'arez and G. Toledo S\'anchez  Phys. Rev. C {\bf 92}, 065204(2015)

\bibitem{qqlinear}G. S. Bali and K. Schilling, Phys. Rev. D {\bf 47} 661(1993). 
\bibitem{moniz}C.~J.~Horowitz, E.~J.~Moniz and John W. Negele, Phys. Rev. D {\bf31} 1689(1985).
\bibitem{oka}M.~Oka and C.~J.~Horowitz, Phys. Rev. D {\bf31} 2773(1985)
\bibitem{Piekarewicz} C.~J.~Horowitz and J.~Piekarewicz  Nucl.~Phys.~ A {\bf 536} 669(1992);
\bibitem{toledo02} G.~Toledo S\'anchez and J.~Piekarewicz, Phys. Rev. C {\bf 65}, 045208 (2002). 
\bibitem{ayala}A. Ayala, J. Magnin, L. M. Montano and G.Toledo Sanchez  
 Phys. Rev. C {\bf 80}  064905(2009). 
\bibitem{formfactor} Samuel H. Blitz and Richard F. Lebed, Phys. Rev. D {\bf91} 094025(2015). 
\bibitem{bicudo13}Pedro Bicudo and Marc Wagner,  Phys. Rev. D {\bf 87}, 114511 (2013).

 
 
\end{thebibliography}
\end{document}